\documentstyle [12pt,epsf]{article}

\begin{document}
%
% ============================ Definitions =============================
%
\newcommand{\mod}[0]{\mbox{ mod }}
\newcommand{\Abs}[1]{|#1|}
\newcommand{\Tr}[0]{\mbox{Tr}}
\newcommand{\EqRef}[1]{(\ref{eqn:#1})}
\newcommand{\FigRef}[1]{fig.~\ref{fig:#1}}
\newcommand{\Abstract}[1]{\small
   \begin{quote}
      \noindent
      {\bf Abstract - }{#1}
   \end{quote}
    }

\title{Periodic orbit quantization of
the Sinai billiard in the small scatterer limit}

\author{Per Dahlqvist\\
Royal Institute of Technology,\\
S-10044 Stockholm, Sweden\\
and \\
G\'abor Vattay\\
E\"otv\"os University Budapest,\\
Department of Solid State Physics,\\
H-1088 Budapest, M\'uzeum krt. 6-8,
Hungary}
\date{}

\maketitle

\begin{abstract}
We consider the semiclassical quantization of the Sinai billiard
for disk radii $R$ small compared to the wave length $2\pi/k$.
Via the application of the periodic orbit theory of diffraction we derive 
the semiclassical spectral determinant.
The limitations of the derived determinant are studied by comparing it
to the exact KKR determinant, which we generalize here for the $A_1$ subspace.
With the help of the Ewald resummation method developed for the full
KKR determinant we transfer the complex diffractive determinant to a
real form.  The real zeros of the determinant are the quantum
eigenvalues in semiclassical approximation.
The essential parameter is the strength of the scatterer 
$c=J_0(kR)/Y_0(kR)$. Surprisingly, this can take 
any value between $-\infty$ and $+\infty$ within the range of validity of the
diffractive approximation $kR \ll 4$, 
causing strong perturbation in spite of the smallness of the disk.
We study the statistics exhibited by spectra for fixed values of  $c$.
It is Poissonian for $c=\pm \infty$, provided the disk is placed inside
a rectangle whose sides obeys some constraints. 
For $c=0$ we find a good agreement of the level spacing distribution
with GOE, whereas the 
form factor and
two-point correlation function are  similar
but exhibit larger deviations.
By varying the parameter $c$ from $0$ to $\pm \infty$ the
level statistics interpolates smoothly between these limiting cases.
Any transitional level statistics can thus be found
in the spectrum of a Sinai billiard with sufficiently small $R$ 
while we go from the quantum to the semiclassical limit $k\rightarrow\infty$.
\end{abstract}

\section{Introduction}

One of the fundamental questions of quantum chaos and mesoscopic
physics is the relation between the classical motion and the energy 
levels of the quantum counterpart of the system\cite{bohigas,eckhardt,
haake}. Correlation of energy levels in chaotic systems in the
semiclassical limit ($\hbar\rightarrow 0$) are expected to coincide with
those in Gaussian Random Matrix Ensembles (RME) with different
symmetries (GOE, GUE etc.), while generic integrable systems are 
characterized by Poissonian level statistics\cite{berry2}.  
Recently significant progress has been made concerning the
applicability of RME for disordered and chaotic systems\cite{Altshuler}.
In disordered systems the two point correlation function has been
expressed in terms of the spectral determinant of the diffusion 
equation.
These results have been derived with
supersymmetric techniques. 
It has been suggested that this result can
be generalized to chaotic 
systems\cite{Altshuler,Agam}, 
where the spectral determinant of the Perron-Frobenius 
operator plays a similar role. 

Similar results have been derived with the
help of semiclassical Gutzwiller type periodic orbit 
expressions\cite{BK}. This approach relies heavily on the so-called
diagonal approximation for $t<t^*$,
where $t^*$ is of the order of the Heisenberg time (see below),
in order to, in an ingenious way,
make statements about $t>t^*$.
The suggested results of \cite{Altshuler,Agam} then follows after some
further approximations.

One of the conclusions of these studies is that deviations from
RME behavior is expected only when the chaotic mixing is slow
and the characteristic time of mixing $T_m$ is comparable with the 
Heisenberg time. In a usual system the mixing time $T_m$
is a classical quantity characterizing the decay of classical
correlation functions and as such it is of order $\hbar^0$.
The Heisenberg time is the uncertainty time $T_H\sim\hbar/\Delta$
corresponding to the mean level spacing ($\Delta$) .
In the semiclassical limit $\hbar\rightarrow 0$ the mean level spacing
behaves like $\Delta\sim\hbar^d$, where $d$ is the dimension of the
system and the Heisenberg time gets much larger than the mixing time.

Slow mixing can be expected if we perturb an integrable
system with a small point like object, whose classical size
can be neglected. In this case the only source of mixing is the
quantum mechanical diffraction of waves on the small object. 
The diffraction constant governing the speed of the
mixing is proportional with $\hbar^2$ and the resulting mixing
time is proportional with $1/\hbar$. 
In a two dimensional system the level spacing is also proportional with
$\hbar^2$ and the ratio of the mixing time and the Heisenberg time remains 
fixed in the semiclassical limit $\hbar\rightarrow 0$.
This gives the possibility to study systems, where corrections
to RME remain finite. In this paper we study the Sinai billiard 
with a small scatterer, which, as we show here, is a strongly 
diffractive system and its level statistics deviates strongly 
from the RME result.

We are considering the standard Sinai billiard, a unit square 
with a reflective circular disk in the middle. 
We then develop the periodic orbit description of the Sinai billiard with 
small scatterer based on the diffractive extension\cite{vattay1} of
the Gutzwiller trace formula. 
This is done in sections 2 and 3.1.
The diffractive determinant we
derive from the trace formula is directly related
to the KKR determinant of Berry, which is discussed in section 3.2.
We will study a straightforward generalization of Berry's method to the $A_1$
subspace, which exhibit some very interesting features, even in the
diffractive or small radius limit.
In section 4 we
study the level statistics 
exhibited by the zeroes of the diffractive determinant for various values
of the parameter $kR$. We then
change from a square to a suitably chosen rectangle shape,
in order to avoid the non generic degeneracies of the square billiard.

\section{Geometric theory of diffraction}

In this section we review the main steps of the derivation of
the diffractive trace formula and spectral determinant introduced in
Ref.\cite{vattay1}.
The emphasis is on point-like diffractive sources with no angular dependence
in their associated diffraction constants.
These point sources are placed inside a bounded and otherwise non diffractive
system.

\subsection{Green functions}

For the description of diffraction effects in the semiclassical limit
($\hbar\rightarrow 0$) Keller\cite{Keller1} introduced the {\em Geometric
Theory of Diffraction} (GTD). According to this theory, the energy domain
Green function can be divided into a geometric ($G$) and a diffractive ($D$) 
part:
\begin{equation}
G(q,q',E)=G_G(q,q',E)+G_D(q,q',E) \ \ ,
\end{equation}
where the geometric part is the semiclassical Green function. 
The semiclassical Green function is a sum
for all classical trajectories ($j$) with energy $E$ connecting the
starting point $q$ and the end point $q'$
\begin{equation}
G_G(q,q',E)= \sum_{j\in {q\rightarrow q'}} G_{0}^{(j)}(q,q',E) \ \ ,  
\label{eqn:pathsum}
\end{equation}
where $G_{0}^{(j)}(q,q',E)$ is the Van Vleck-Gutzwiller Green function.
It can be calculated from the action 
$S_j(q,q',E)=\int_q^{q'} pdq$ calculated along
trajectory $j$
\begin{equation}
G_0^{(j)}(q,q',E)=A_j(q,q',E)e^{\frac{i}{\hbar}S_j(q,q',E)} \ \ ,
\end{equation}
where the amplitude $A_j(q,q',E)$
%$$A_j(q,q',E)=\frac{1}{\hbar^2}\left|\det\frac{\partial^2 S_j(q,q',E)}{\partial
%q_i\partial q_j'}\right|^{1/2} e^{-i\pi\nu_j/2}$$
%\footnote{check!} 
is the Van Vleck-Gutzwiller
determinant \cite{Gut3} multiplied with the Maslov phase.

The diffractive part of the semiclassical Green function $G_D(q,q',E)$
describes the effect of wave scattering on points
$q^s_k$ being singular points of the classical
dynamics\cite{vattay1,niall}. It is a sum
over all possible trajectories
which start from $q$, go to one of the singular points $q^s_k$, then
leave the singular point and reach the end point $q'$; or go
and visit any combination of singular points before reaching the
end point.  
A trajectory going trough one singularity 
contributes to 
the Green function with\cite{Keller1}
\begin{equation}
G_{0}^{(j)}(q,q^s_i,E) d_k(p,p',E) G_{0}^{(j')}(q^s_i,q',E) \ \ . \label{eqn:onescatt}
\end{equation}
The generalization to multi scattered trajectories is straightforward,
and will be developed in the
next section.
In eq \EqRef{onescatt},
$j$ and $j'$ are indices of the incoming and
the outgoing classical trajectory, $d_k(p,p',E)$ is the diffraction constant 
associated with
the singularity located at $q^s_k$. It depends on the direction of incoming and
outgoing momenta $p$ and $p'$ and energy $E$. 
The diffraction constant is the leading
part of the exact quantum mechanical scattering amplitude of the
singularity.

\subsection{Traces and periodic orbits}

Based on the expression for the Green functions in the previous subsection
we can derive the
diffractive version of the Gutzwiller trace formula\cite{vattay1}.

The trace of the total Green function is a sum of its geometric
and diffractive parts:
\begin{equation}
Tr G (E)= Tr \tilde{G}_G (E)+ Tr G_D(E)+ g_0(E)\ \ \label{sm} ,
\end{equation}
where the geometric part $Tr \tilde{G}_G(E)$ is just the usual Gutzwiller trace
formula, a sum over unstable periodic orbits
possibly amended by a sum over neutral orbits and
the contribution from zero length orbits is in $g_0(E)$.
$\tilde{G}_G(0,0,E)$ is essentially the Green function
$G_G(0,0,E)$ without the singular contribution coming from the zero 
length orbit.
We concentrate here on the new term
\begin{equation}
Tr\; G_D (E)=\int G_D(q,q,E)dq \ \ .
\end{equation}

To be more specific, we assume that the diffraction is caused by
the presence of
$N$ small disks
located at $q^s_k$, where $1\leq k \leq N$, whose diffraction constants
$d_k(E)$ do not depend on the scattering angle\cite{wirzba}.
This will simplify the formulation considerably.
We can easily see that the
contribution to the Green function from all orbits
via the disks $k_1$ and $k_2$
\begin{equation}
\sum_{j_0;q \mapsto q_{k_1}}
G_0^{(j_0)}(q,q_{k_1}^s)d_{k_1}
\sum_{j_1;q_{k_1} \mapsto q_{k_2}}
G_0^{(j_1)}(q_{k_1}^s,q_{k_2}^s)
d_{k_2} ...\sum_{j_2:q_{k_2} \mapsto q'} 
G_0^{(j_2)}(q_{k_2}^s,q')  \label{eqn:Green_2}
\end{equation}
\[
=\sum_{j_0;q \mapsto q_{k_1}}
G_0^{(j_0)}(q,q_{k_1}^s)d_{k_1}
\; \; \tilde{G}_G(q_{k_1}^s,q_{k_2}^s)
d_{k_2} ... \sum_{j_2:q_{k_2} \mapsto q'} 
G_0^{(j_2)}(q_{k_2}^s,q') \ \ .  
\]
The generalization to multi scattered trajectories is now
obvious.

We can introduce a symbolic dynamics by simply labeling the disks by the
corresponding indices $k$. 
In an open N-disk system a periodic disk sequence 
$\overline{k_1 k_2 \ldots k_n}$ codes the single
periodic orbit visiting the disks in the prescribed order.
In a bounded system it denotes {\em all} such periodic
orbits e.g. the transition $\ldots k_i k_{i+1} \ldots$ denotes all
(topologically distinct) ways to go from disk $k_i$ to disk $k_{i+1}$.
%via an arbitrary number of bounces on the billiard walls.
In the following we will refer to {\em periodic symbol codes} and
{\em symbolic periodic orbits} to avoid confusion with periodic orbits.

The calculation of the trace of the Green function $G_D(q,q',E)$ 
amounts to add up 
contributions from all {\em distinct}
paths from $q$ to $q'$ via $n\geq 1$ disk scatterings and then 
taking the trace.
The basic building block $\tilde{G}_G(k_i,k_{i+1},E)$ in semiclassical
approximation is
a sum over all paths from (see \EqRef{pathsum}).

Consider now a situation where the starting and ending points  coincide
$q=q'$. 
The contribution
from the symbolic path $k_1\;, k_2 \ldots k_n$ to the trace is
\begin{equation}
\int dq \sum_{j_0;q \mapsto q_{k_1}}
G_0^{(j_0)}(q,q_{k_1}^s)d_{k_1}\tilde{G}_G(q_{k_1}^s,q_{k_2}^s)
d_{k_2}\cdots 
\tilde{G}_G(q_{k_{n-1}}^s,q_{k_n}^s)d_{k_n} \sum_{j_n:q_{k_n} \mapsto q} 
G_0^{(j_n)}(q_{k_n}^s,q) \ \ . \label{eqn:Green_n}
\end{equation}
The integral over $q$ can be performed  in the $\hbar\rightarrow 0$ limit
by using the saddle point
method. 
The result is
\[
\frac{1}{i\hbar}\sum_{j:q_{k_n} \mapsto q_{k_1}} 
\left(T^{(j)}(q_{n}^s,q_{1}^s) 
G_0^{(j)}(q_{n}^s,q_{1}^s)  \right) 
\]
\begin{equation}
\cdot 
\tilde{G}_G(q_{q_{k_1}}^s,q_{q_{k_2}}^s) d_{k_1} 
\tilde{G}_G(q_{k_2}^s,q_{k_3}^s) d_{k_2}\cdots 
\tilde{G}_G(q_{q_{n-1}}^s,q_{k_
n}^s)d_{k_n} \ \ , \label{eqn:trace_n}
\end{equation}
The stationary phase conditions select out all points $q$ lying on a trajectory
between disk $k_n$ and $k_1$, that is, it selects periodic orbits.

To get the full contribution from the symbolic periodic orbit
$\overline{k_1k_2\ldots k_n}$
one must add the contribution from all {\em nonidentical}
cyclic permutations of the sequence $k_1k_2\ldots k_n$
in eqs. \EqRef{Green_n} and \EqRef{trace_n}.
The symbolic periodic orbit $\overline{k_1 k_2\ldots k_n}$
can be a repetition of a shorter {\em primitive} (symbolic) periodic orbit
$\overline{k_1k_2\ldots k_n}=\overline{p^r}$. 
The length of the code $p$ is denoted
$n_p=n/r$.
Instead of restricting ourself to all non identical 
permutations we can take all
cyclic permutations and then divide by $r$.

The result of this somewhat laborious exercise in
combinatorics can be
conveniently expressed 
in terms of the derivative of the
quantity
\begin{equation}
t_p=\prod_{j=1}^{n_p} d_{k_j} 
\tilde{G}_G(q_{k_j},q_{k_{j+1}})
\end{equation}
as
\begin{equation}
Tr G_D=\sum_{p \in \Omega_D}  
\sum_{r=1}^{\infty} \frac{1}{r} \frac{d}{dE} t_p^r
\end{equation}
where $\Omega_D$ is the set of all primitive 
periodic symbol sequences taken from the alphabet  $\{ k; 1\leq k \leq N \}$ where
$N$ is the number of singular points in the system.

We now introduce the spectral determinant $\Delta(E)$, defined through
the logarithmic derivative
\begin{equation}
Tr G=-\frac{d}{dE} \log \Delta(E) \ \ .
\end{equation}
Since the Green function is a sum (see Eq.\ref{sm}),
the spectral determinant naturally factorizes as a product
\begin{equation}
\Delta(E)=\Delta_0(E)\cdot\Delta_G(E)\cdot\Delta_D(E) \ \ ,
\end{equation}
where $\Delta_0(E)$ is associated with zero length orbits and corresponds 
to the mean level density, $\Delta_G(E)$ is the Gutzwiller-Voros zeta 
function associated with classical periodic orbits,
possibly amended by neutral orbits.
The diffractive determinant (or zeta function) now reads
\begin{equation}
\Delta_D(E)=\prod_{p \in \Omega_D} (1-t_p) \ \ \label{spede}.
\end{equation}

\section{Application to the Sinai billiard}

In this section we work out the diffractive spectral
determinant of the Sinai billiard in the small
disk limit. Then we show that the diffractive
determinant is essentially the leading part of the 
Korringa-Kohn-Rostoker (KKR) determinant in the small disk limit.

\subsection{The diffractive spectral determinant}

The application for the Sinai billiard is now very simple, provided
$kR$ is sufficiently small.
The diffraction constant has been derived in Ref. \cite{wirzba}, 
in the limit $kR \rightarrow 0$ it was  found to be
\begin{equation}
d(E)\sim -4i\frac{J_0(kR)}{H_0^{(1)}(kR)}  \  \ .  \label{eqn:d}
\end{equation}
It does not depend on the opening angle
and the considerations in the previous section can be applied.

The billiard contains only a single disk and the symbolic dynamics introduced
in the previous section has only a single letter 
implying the existence of one single primitive periodic code,
and the resulting determinant
is trivial
\begin{equation}
\Delta_D(E)=1-d(E)\tilde{G}_G(0,0,E) \ \ .
\end{equation}

The geometry of the Sinai billiard is easy to visualize
if one unfolds it into a quadratic lattice of disks.
A trajectory from the diffractive object (i.e. the small disk)
and back to itself is characterized by
a lattice vector $\rho=(\rho_x,\rho_y)$ where
$\rho_x$ and $\rho_y$ can take any integer value. 
The geometric Green function $G_G(0,0,E)$ may be expressed as a sum
over lattice vectors
\begin{equation}
\tilde{G}_G(0,0,E)=\sum_{\rho \neq (0,0)} G_0(0,\rho,E)\ \ ,
\end{equation}
with the zero length vector $\rho = (0,0)$ excluded.
The Green function $G_0$ may be expressed in terms of (the semiclassical limit of)
a free Green function
\begin{equation}
\tilde{G}_G(0,0,E)=\sum_{\rho \neq (0,0)} G_{free}(0,\rho,E) (-1)^{\rho_x+\rho_y}\ \ ,
\end{equation}
where
$(-1)^{\rho_x+\rho_y}$ is a phase index associated with the reflection on
the square wall.
But in order to improve the
approximation we
use the exact free Green function in two dimensions
\begin{equation}
G_{free}(x,x')=-\frac{i}{4}H_0^{(1)}(k|x-x'|)  \ \ .
\end{equation}
The resulting expression is
\begin{equation}
\Delta_D(E)=1-d(E)\sum_{\rho\neq (0,0)} (-\frac{i}{4})H_0^{(1)}(k|\rho|)
(-1)^{\rho_x+\rho_y}  \label{eqn:Zcyc} \ \ ,
\end{equation}
where $k=\sqrt{2E}$ is the momentum in dimensionless units
$\hbar=m=1$. 
The sum for the lattice vectors is the exact Green function 
of the empty square with the zero length term removed.

The Green function can also be expressed via the eigenfunctions of
the empty square. The wave functions can be categorized
according to the different
irreducible representations of $C_{4v}$. There are four one dimensional 
($A_1$, $A_2$, $B_1$ and $B_2$) representations
and one two dimensional ($E$). 
The disk is centered at the point where all the symmetry lines meet.
Wave functions of the empty square billiard vanish at this point except
for the $A_1$ subspace. Accordingly, only the energies belonging to
this subspace are affected by 
diffraction in leading semiclassical order as will be demonstrated
later.

\subsection{The KKR determinant}

The full quantum
Sinai billiard problem can be solved with the
KKR method introduced in this context by Berry in Ref.\cite{berry2},
where the $A_2$ subspace has been worked out in detail.
The method leads to a determinant $D_{KKR}$ whose
zeroes yield the eigenvalues of the problem.
The generalization to other subspaces are fairly straightforward.
Here, without repeating the derivation of Ref.\cite{berry2},
we quote the relevant results for the $A_1$
and $A_2$ subspaces and show that the periodic orbit expression 
\EqRef{Zcyc} is essentially the leading part of the KKR
determinant.

The effective dimension of the KKR matrix is $\sim (kR)/4$.
For $kR<4$, which we call the diffractive zone, it is essentially 
one dimensional. The determinant is given by the leading element of 
the matrix
\begin{equation}
\Delta_{KKR}(k)= \left\{ \begin{array}{ll}
1+\frac{J_0(kR)}{Y_0(kR)}S_0^r(e) & \mbox{ in the $A_1$ subspace} \\
1+\frac{J_4(kR)}{Y_4(kR)}(S_0^r(e)-S_8^r(e)) & \mbox{in the $A_2$
subspace}  \end{array}
\right.  \  \  ,
\label{eqn:DA1A2}
\end{equation}
where the real functions $S_l^{(r)}(e)$ are related to the {\em structure constants} 
\begin{equation}
S_l(e)=-i \sum_{{\bf \rho}\neq{\bf 0}} H_l^{(1)}(k\rho)e^{il\phi_{\bf \rho}}
\cdot \left\{ \begin{array}{ll}
(-1)^{\rho_x+\rho_y} & \mbox{ in the $A_1$ subspace} \\ 
1 &  \mbox{in the $A_2$ subspace} \\
\end{array} \right. \ \ . \label{eqn:Sl}
\end{equation}
via
\begin{equation}
S_l(e)=S_l^r(e)+i\delta_{0,l}  \  \ .
\end{equation}
$e=(k/2\pi )^2$ is a rescaled energy (denoted by $E$ in \cite{berry2}).

The structure constants, as they stand, do not converge.
They can be evaluated by transforming them to sums over the dual lattice
using the
Ewald resummation method\cite{berry2}.
The resulting expressions are
\begin{eqnarray}
S_l^{(r)}(e)\approx\frac{1}{\pi^2} \sum_{{\bf v}} (\frac{\mbox{v}^2}{e})^{l/2}
\frac{\exp (l/2[ 1-\mbox{v}^2/e])}{e-\mbox{v}^2}  &  l\neq 0\\
S_0^{(r)}(e)\approx \frac{1}{\pi^2} \sum_{{\bf v}} 
\frac{\exp (Q[ 1-\mbox{v}^2/e])}{e-\mbox{v}^2}-E_i(Q)/\pi & l=0
\end{eqnarray}
We have introduced the vector ${\bf v}$ in the dual lattice
\begin{equation}
\begin{array}{ll}
{\bf v}=(v_x+\frac{1}{2},v_y+\frac{1}{2}) & \mbox{ in the
$A_1$ subspace}\\
{\bf v}=(v_x,v_y) & \mbox{ in the $A_2$ subspace}  \ \ . \\
\end{array}
\end{equation}
where $v_x$ and $v_y$ are integers.
The poles of the expressions located in $\mbox{v}^2$ are the eigenvalues of
the empty square billiard.
The smaller the number $Q$ the more accurate the method works.

Combining Eqs. \EqRef{d}, \EqRef{Zcyc}, \EqRef{DA1A2} and \EqRef{Sl}
we find the desired relation between the diffractive limit of the KKR 
determinant and the diffractive determinant
\begin{equation}
\Delta_{KKR}(k)=\left(1-i\frac{J_0(kR)}{Y_0(kR)}\right)\Delta_D(k) \ \ .
\end{equation}
This relation is one of the  main results of the paper.

The KKR determinant, 
and thus the diffractive spectral determinant,
exhibit poles at the unperturbed eigenvalues $e_{\bf v}^{(0)}=\mbox{\bf v}^2$.
They cancel the zeroes of the geometric determinant $\Delta_G(k)$
according to the formula
\begin{equation}
\Delta(k) \sim \Delta_G (k) \Delta_D(k) \ \ ,
\end{equation}
and give rise to corrected eigenvalues given by the zeroes of $\Delta_D(k)$.
This is an important property of diffractive zeta functions:
they cancel the geometric zeros and create new ones.

Note that there is no contribution coming from isolated unstable orbits
in our system,
$\frac{d}{dE}\log \Delta_G(k)$ is just the Berry-Tabor sum for the square
billiard \cite{BT}.

\section{Investigation of the energy levels}

The investigation of the KKR determinant for all symmetry subspaces
is not a superfluous exercise. Next we show that the
$A_1$ subspace will exhibit some very interesting features, 
to which we devote the rest of the paper.

The quantum eigenvalues 
in the diffractive limit
can be determined from the equation
\begin{equation}
\Delta_{D}(k)=0.
\end{equation}
or euivalently from the real equation
\begin{equation}
S^{(r)}(e)+c(kR)=0 \label{eqn:S0c} \ \ ,
\end{equation}
where
\begin{equation}
c(kR)=\frac{Y_0(kR)}{J_0(kR)} \ \ .
\end{equation}

\subsection{Perturbation theory}

When $c(kR)$ is large one can use 
perturbation theory
to compute the roots of Eq. \EqRef{S0c} (see  Ref.\cite{berry2}).
One thus
assumes that the difference between a perturbed level $e$ and an
unperturbed level $e^{(0)}$ is small.
$S_0^{(r)}(e)$ close to $e^{(0)}$ may then be approximated by 
\begin{equation}
S_0^{(r)} \approx \frac{8}{\pi^2(e-e^{(0)})} \ \ ,
\end{equation}
yielding the perturbed eigenvalue
\begin{equation}
e\approx e^{(0)}-\frac{8}{\pi^2}\frac{J_0(kR)}{Y_0(kR)}  \ \ .
\end{equation}
Fig. 1 illustrates how well this expression works for $R=0.025$,
where we have plotted the distance from an eigenvalue of the Sinai
billiard ($R=0.025$) and  the nearest eigenvalue of the square billiard.
The expression above applies only to non-degenerate states.
For a $n$-degenerate unperturbed energy,
$n-1$ levels remain almost unaffected, which is clearly seen on Fig. 1.

There are two regions in the range $0< kR <4$ where perturbation
theory should work in principle.

The first is the limit $kR \rightarrow 0$.
Taking well known asymptotic expressions\cite{AS} 
for the Bessel functions one gets
\begin{equation}
e=e^{(0)}-\frac{1}{2\pi}\frac{1}{\log (kR/2) + \gamma}\ \ ,
\end{equation}
where $\gamma=0.57721 \ldots$ is Euler's constant.
However, this expression goes very slowly 
to zero as $kR \rightarrow 0$ and unless
the radius is very small perturbation theory 
does not even apply to the ground state
$e_1^{(0)}=0.5$.

The second is where $c(kR)$
has a pole within the diffractive zone $kR<4$, namely at
$kR=j_{0,1}=2.40482\ldots$.
Close to this value, the perturbation expression works
very well, as it can be seen on the scatter plot Fig. 1.
Note, that the index of the levels is now shifted one step up relative
to the unperturbed ones, this is why we chose to plot the nearest distance to 
an unperturbed eigenvalue.
We note that $c(kR)$ 
exhibit a zero at $kR = y_{0,1}\approx 0.899$ so the perturbation theory
also breaks down deep inside the diffractive zone.

As a comparison,
in the $A_2$ subspace the perturbation is on average\cite{berry2}
\begin{equation}
<e-e_0>=
-\frac{8}{\pi^2}\frac{J_4(kR)}{Y_4(kR)}\sim
-\frac{1}{18 \pi}\left( \frac{kR}{2}\right)^8  \ \ .
\end{equation}
Perturbation theory works well almost throughout the whole diffractive region.
So the spectrum in the $A_2$ representation, in the diffractive zone,
can hardly be expected to be particularly interesting.

For $A_1$, on the other hand, $|c(kR)|$ will range from $0$ to $\infty$
in the interval $0<kR<y_{0,1}\approx 0.899$ which is well inside the
diffractive zone.

\subsection{Level statistics in the diffractive limit}

In this section we solve Eq. \EqRef{S0c} 
numerically and study the statistics of levels.
The essential parameter of the problem is $kR$ since the mean level
spacing $\Delta_e$ does not depend on energy $e$. 
Here we will study spectra for fixed values
of $c(kR)$. 
The idea of keeping $c$ and thus $kR$ fixed  needs some explanation.
If we decrease $R$ and keep $kR$ fixed by increasing $k$, more and more
levels are squeezed in a neighborhood of some
$k =k_0$. We then can safely
replace $kR$ by $k_0 R$ and it is possible to obtain
a sufficiently large number of levels around $k_0$ for which
the value of $c$ is approximately
constant and given by $c(k_0R)$.
For obvious reasons we calculate the bottom part of the spectrum.
However, this sample should exhibit the same statistics as any comparable
sequence of levels, e.g. around $k_0$. 
The reason is that the location of a level depends
on the parameter $kR$ and the location of the unperturbed levels
in the environment and the statistics of these unperturbed levels is
well defined and given by a Poissonian process.

In the actual calculation we made a minor modification of the
conventional Sinai billiard, since 
the empty square billiard has degeneracies in its spectrum.
%a highly degenerate spectrum.
Their origin is number theoretical and is connected with the high
degree of symmetry in the system. If a disk is present,
these degeneracies are split
by higher order terms in the KKR determinant. 
Within the scope of the present diffractive determinant they remain
there.
In the following we want to avoid this complication and replace the
square by a rectangle with sides $a=1$ and $b=2^{1/4}$.
The function $S_0^{(r)}(e)$ is modified accordingly 
\begin{equation}
S_0^{(r)}(e)\approx \frac{1}{ab}\frac{1}{\pi^2} \sum_{{\bf v}} 
\frac{\exp (Q[ 1-\mbox{v}^2/e])}{e-\mbox{v}^2}-E_i(Q)/\pi,
\end{equation}
where ${\bf v}=((v_x+1/2)/a,(v_y+1/2)/b)$ is now the dual lattice of the rectangle.

We investigate three statistical measures on the spectra.
First we will investigate the integrated probability distribution of 
neighboring levels 
\begin{equation}
P(s)=\sum_n\theta(s-(e_{n+1}-e_n)/\Delta_e) \ \ ,
\end{equation}
where $\Delta_e$ is the mean level spacing around the energy $e$ and
$\theta(x)$ is the step function.
Then we investigate the two point correlation function of levels
\begin{equation}
R(\epsilon)=
\langle \;
\sum_{ij} \delta((e-e_i)/\Delta_e+\epsilon/2)\cdot
\delta((e-e_j)/\Delta_e-\epsilon/2) \;
\rangle_e
\end{equation}
\[
=\delta(\epsilon)+\langle \;
\sum_{i\neq j} \delta((e-e_i)/\Delta_e+\epsilon/2)\cdot
\delta((e-e_j)/\Delta_e-\epsilon/2) \;
\rangle_e
\equiv
\delta(\epsilon)+\tilde{R}(\epsilon)
\]
where the average is taken for a large number of energies.
We also compute its Fourier transform, the form factor,
\begin{equation}
K(\tau)=\int_{-\infty}^{+\infty}e^{2\pi i \tau \epsilon}(R(\epsilon)-1)d\epsilon
\end{equation}
The correlation function and the form factor are computed over a gaussian window
centered at the middle of the sample spectrum, its width is about  one sixth of the
sample size.
The results are then smeared with another gaussian.
We choose the width of this gaussian 
rather small so some (non significant) oscillations due to finite statistics
will remain
in the plots.

In Fig. 2 we show results for $c=0$,
with 5000 levels computed.
The level spacing curve (Fig 2a) is close to the prediction
of the random matrix theory (GOE), with a small but
significant deviation.
An exact agreement is hardly possible because
the levels are locked
between the unperturbed levels\cite{berry2}.
The close agreement with GOE is in this perspective
rather surprising.

The formfactor $K(\tau)$ is plotted in Fig 2b.
There is again good agreement with GOE except for
small $\tau$.
The bump for small values of $\tau$ 
indeed persist rather high up in the spectrum (of the full
KKR determinant)
and is quite similar to the behavior predicted in \cite{PDLA}.

The correlation function $\tilde{R}(\epsilon)$ is plotted in Fig. 2c.
The deviation from GOE is now most pronounced around $\epsilon\approx 1$.

In Fig. 3 we show  
the correlation function $\tilde{R}(\epsilon)$
for a geometric sequence of $c$'s,
each spectrum contains 1500 levels.
We note how it, in a uniform way, interpolates between a Poissonian
result and the GOE-like. 

\section{Concluding Remarks}

The KKR determinant may be recast into a periodic orbit 
expressions
in the limit $kR \rightarrow \infty$ \cite{berry2}.
Periodic orbit calculations
are very laborious in this limit, it is questionable
whether it provides a useful method of extracting (many) eigenvalues for bound
chaotic systems\cite{PDNaka}.

What motivated us in this paper, 
is that the KKR determinant
yields (diffractive) periodic orbit expressions also in the limit 
$kR \rightarrow 0$.
This diffractive limit is very feasible form a periodic orbit point of view.

A traditional symbolic dynamics for the Sinai billiard uses the lattice vectors
$\rho$ as symbols \cite{PDsin}. 
In the diffractive limit, this symbolic dynamics
is
complete.
The reason for this is the absence of shadow behind disks in the 
$kR \rightarrow 0$ limit. Moreover, the diffraction constant
is independent of the scattering angle.
In the language of cycle expansions\cite{AAC}, 
these circumstances imply that
the cycle expansion \EqRef{Zcyc} is entirely
given by the (well defined) fundamental part, the curvature correction vanish
identically.

So, what we have presented here is a very straight forward application of
the cycle expansion technique as well as of
the geometric theory of diffraction. Yet,
the results are non-trivial.
Even if we have not used the diffractive determinant for computations 
as it stands \EqRef{Zcyc}
we have demonstrated that study of spectral statistics is in principle
approachable for periodic orbit theories in this case.

A natural extension of this work is to increase the number of 
small scatterers
and study whether GOE emerges as the number of scatterers tends to infinity
and are distributed randomly,
preliminary results indicate that this is indeed
the result \cite{disorder}.

\vspace{0.3cm}

This work was supported by the Swedish Natural Science
Research Council (NFR) under contract no. F-AA/FU 06420-312
and no. F-AA/FU 06420-311., the
Hungarian Science Foundation OTKA (F019266/F17166/T17493) and the Hungarian
Ministry of Culture and Education FKFP 0159/1997.
G. V. thanks for the hospitality of the Mechanics
Department of the Royal Institute of Technology during his stay.

\newcommand{\PR}[1]{{Phys.\ Rep.}\/ {\bf #1}}
\newcommand{\PRL}[1]{{Phys.\ Rev.\ Lett.}\/ {\bf #1}}
\newcommand{\PRA}[1]{{Phys.\ Rev.\ A}\/ {\bf #1}}
\newcommand{\PRD}[1]{{Phys.\ Rev.\ D}\/ {\bf #1}}
\newcommand{\PRE}[1]{{Phys.\ Rev.\ E}\/ {\bf #1}}
\newcommand{\JPA}[1]{{J.\ Phys.\ A}\/ {\bf #1}}
\newcommand{\JPB}[1]{{J.\ Phys.\ B}\/ {\bf #1}}
\newcommand{\JCP}[1]{{J.\ Chem.\ Phys.}\/ {\bf #1}}
\newcommand{\JPC}[1]{{J.\ Phys.\ Chem.}\/ {\bf #1}}
\newcommand{\JMP}[1]{{J.\ Math.\ Phys.}\/ {\bf #1}}
\newcommand{\JSP}[1]{{J.\ Stat.\ Phys.}\/ {\bf #1}}
\newcommand{\AP}[1]{{Ann.\ Phys.}\/ {\bf #1}}
\newcommand{\PLB}[1]{{Phys.\ Lett.\ B}\/ {\bf #1}}
\newcommand{\PLA}[1]{{Phys.\ Lett.\ A}\/ {\bf #1}}
\newcommand{\PD}[1]{{Physica D}\/ {\bf #1}}
\newcommand{\NPB}[1]{{Nucl.\ Phys.\ B}\/ {\bf #1}}
\newcommand{\INCB}[1]{{Il Nuov.\ Cim.\ B}\/ {\bf #1}}
\newcommand{\JETP}[1]{{Sov.\ Phys.\ JETP}\/ {\bf #1}}
\newcommand{\JETPL}[1]{{JETP Lett.\ }\/ {\bf #1}}
\newcommand{\RMS}[1]{{Russ.\ Math.\ Surv.}\/ {\bf #1}}
\newcommand{\USSR}[1]{{Math.\ USSR.\ Sb.}\/ {\bf #1}}
\newcommand{\PST}[1]{{Phys.\ Scripta T}\/ {\bf #1}}
\newcommand{\CM}[1]{{Cont.\ Math.}\/ {\bf #1}}
\newcommand{\JMPA}[1]{{J.\ Math.\ Pure Appl.}\/ {\bf #1}}
\newcommand{\CMP}[1]{{Comm.\ Math.\ Phys.}\/ {\bf #1}}
\newcommand{\PRS}[1]{{Proc.\ R.\ Soc. Lond.\ A}\/ {\bf #1}}

%\end{document}
\newpage

\begin{figure}
\epsffile{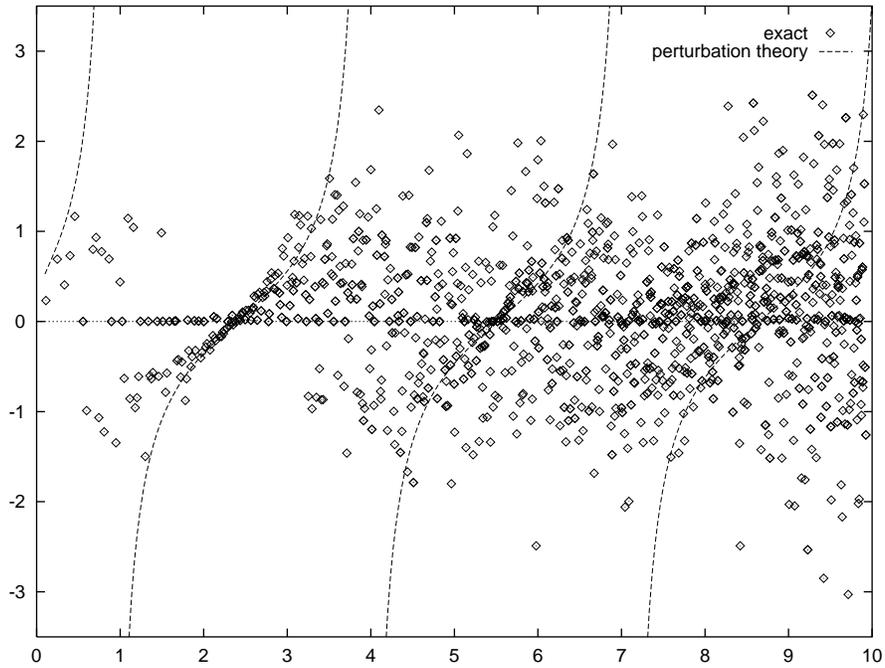}
\caption{Distance between eigenvalues of the Sinai billiard and the
nearest eignvalue of the empty square compared with perturbation theory.
The eigenvalues are from the $A_1$ subspace for radius $R=0.025$.}
\end{figure}

\begin{figure*}
\epsffile{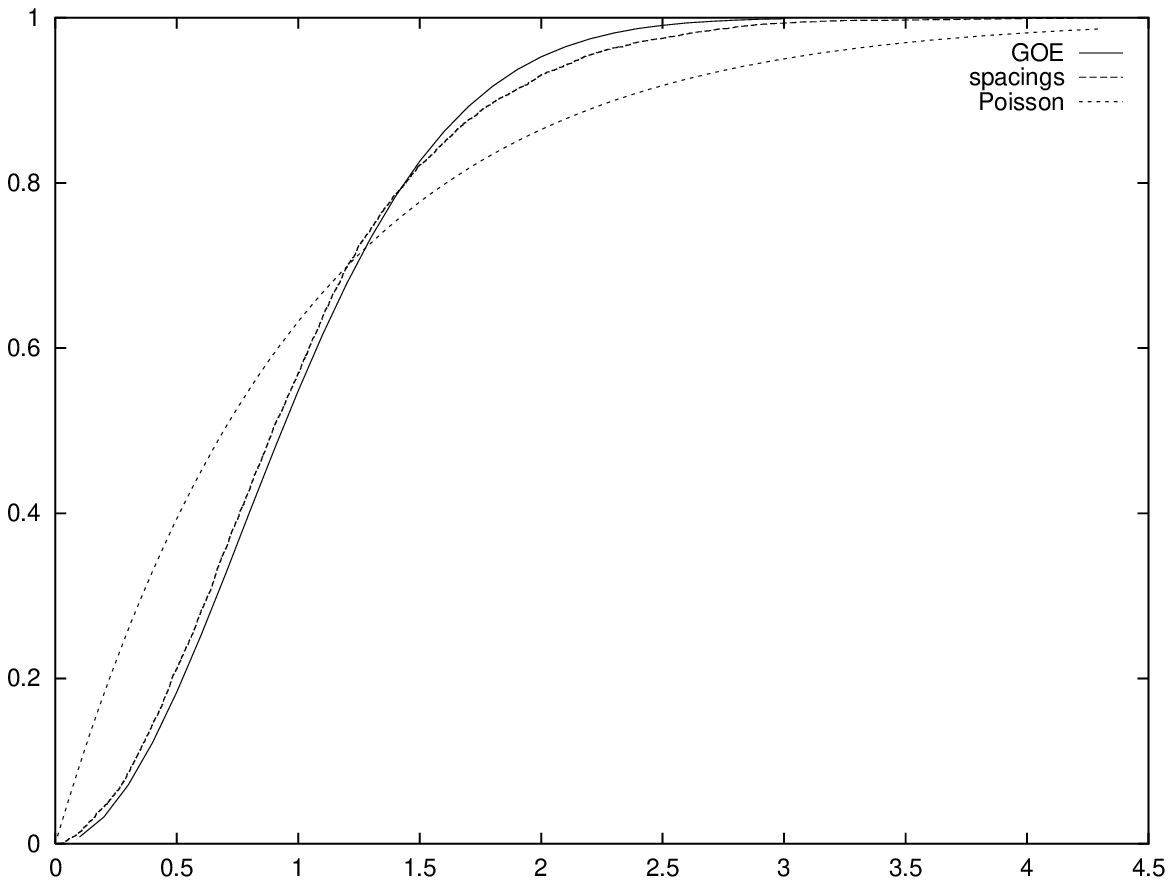}
\epsffile{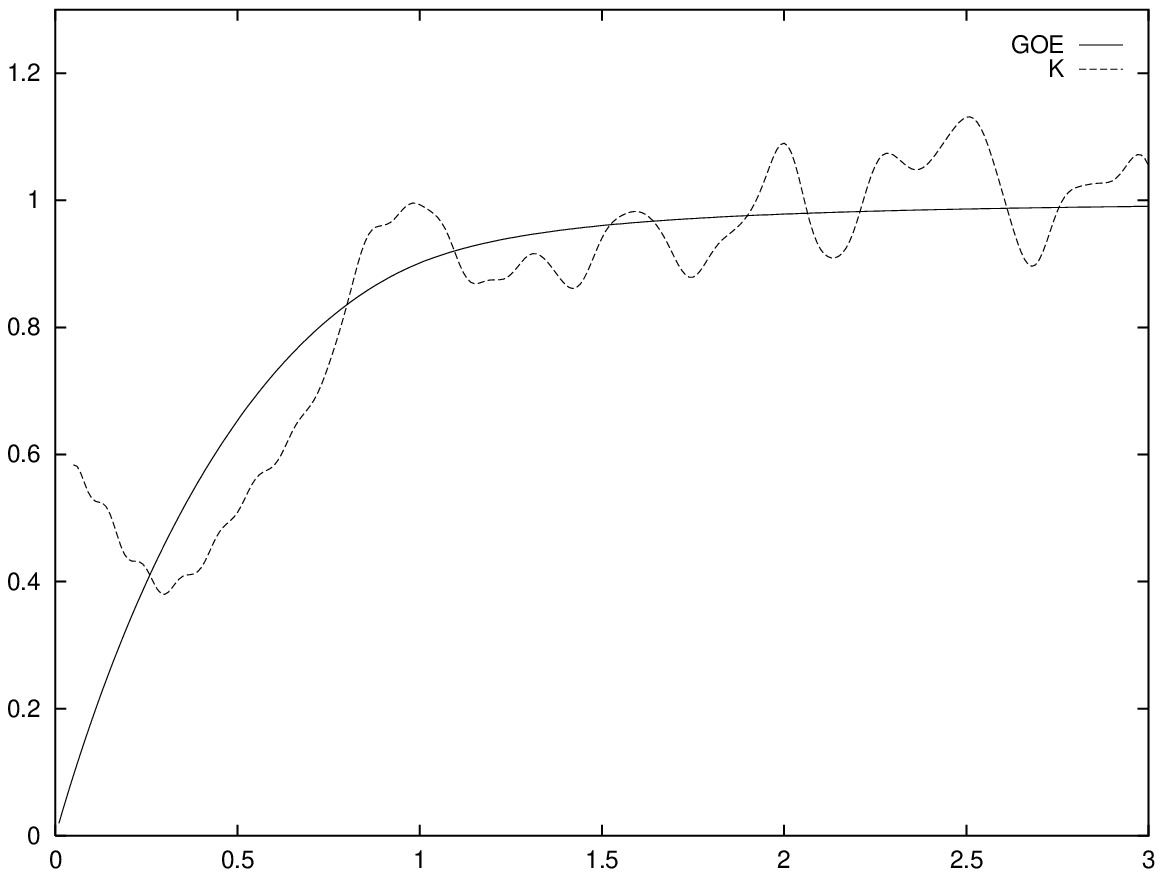}
\end{figure*}

\begin{figure}
\epsffile{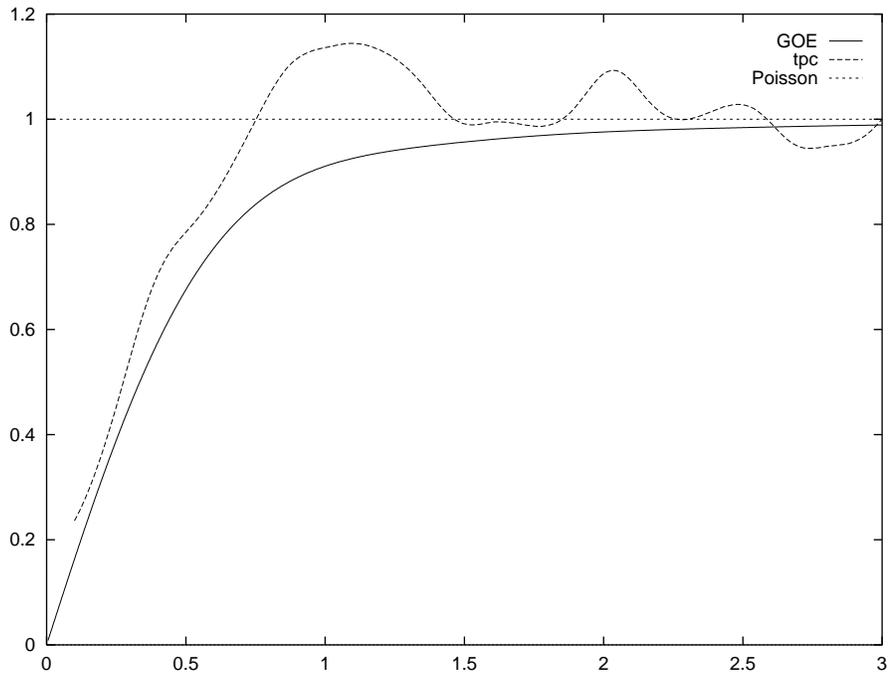}
\caption{
a) Integrated level spacing distribution. b) Form factor. c) Two point
correlation function.
Eigenvalues computed from eq \EqRef{S0c} 
with $c=0$ for a rectangular Sinai billiard.}
\end{figure}

\begin{figure}
\epsffile{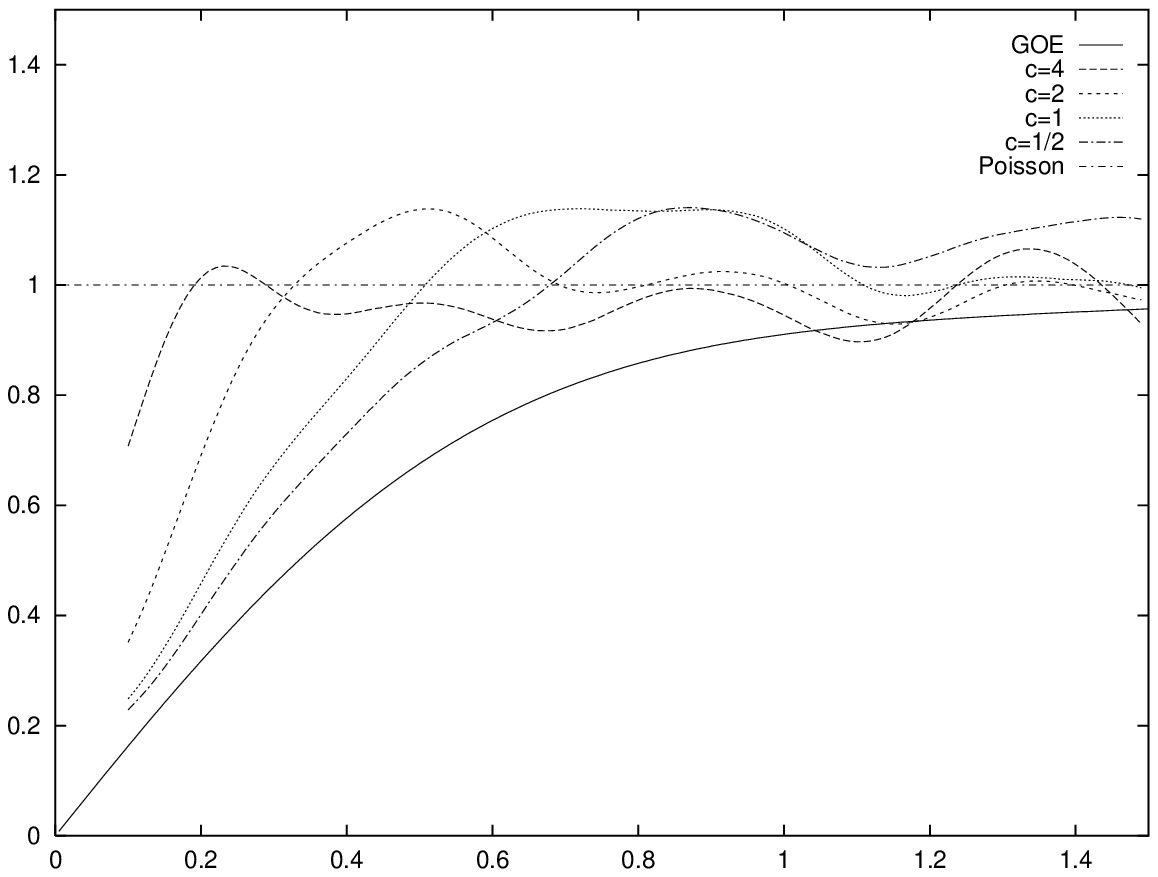}
\caption{Two point correlation function from eq \EqRef{S0c}
for different values of $c$.}
\end{figure}


\begin{thebibliography}{99}
\bibitem{bohigas} O. Bohigas and M.-J. Giannoni, in 
{\em Mathematical and Computational
Methods in Nuclear Physics}, ed.: J. Desha, J. M. Gomez and A. polls, Lecture
Notes in Physics, Vol. 209 (Springer, Berlin 1984)
\bibitem{eckhardt} B. Eckhardt, Phys. Rep. {\bf 163}, 205 (1988)
\bibitem{haake} F. Haake, {\em Quantum Signatures of Chaos} (Springer, Berlin 1991)
\bibitem{Altshuler} A. V. Andreev and B. L. Altshuler, Phys. Rev. Lett.
{\bf 75}, 902 (1995)
\bibitem{Agam} O.~Agam, A.~V.~Andreev and B.~D.~Simons,
 Chaos Solitons and Fractals,
{\bf 8}, 1099, (1997).
\bibitem{BK} E.~B.~Bogomolny and J.~P.~Keating, \PRL{77}, 1472 (1996).
\bibitem{vattay1} G. Vattay, A. Wirzba and P. E. Rosenqvist, Phys. Rev. Lett. {\bf 73},
2304 (1994)
\bibitem{Keller1} J. B. Keller, J. Opt. Soc. Amer. {\bf 52} 116 (1962)
\bibitem{Gut3} M. C. Gutzwiller, {\em Chaos in classical and quantum
mechanics},
Springer Verlag, New York (1990)
\bibitem{niall} H. Bruus and N. Whelan, Nonlinearity   {\bf 9}, 1023-1047 (1996)
\bibitem{wirzba} P. Rosenqvist, N. D. Whelan, A. Wirzba,
J. Phys. {\bf A29}, 5441 (1996)
\bibitem{berry2} M. V. Berry, Ann. Phys. N.Y. {\bf 131}, 163 (1981)
\bibitem{BT} M.~V.~Berry and M.~Tabor, Proc.\ R.\ Soc.\ Lond.\ A {\bf 349},
 101 (1974).
\bibitem{AS}    M.~Abramovitz and I.~A.~Stegun, {\em Handbook of
                mathematical functions}, Washington: National
                Bureau of Standards, (1964).
\bibitem{PDLA}  P.~Dahlqvist, \PD{83}, 124 (1995).
\bibitem{PDNaka} P.~Dahlqvist, Chaos Solitons and Fractals,
{\bf 8}, 1011, (1997).
\bibitem{PDsin} P.~Dahlqvist, Nonlinearity {\bf 8}, 11 (1995).
\bibitem{AAC}   R.~Artuso, E.~Aurell and P.~Cvitanovi\'{c},
                Nonlinearity {\bf 3}, 325 and 361, (1990).
\bibitem{disorder} P.~Dahlqvist, {\em From chaotic to disordered
systems - a periodic orbit approach}, in preparation.

\end{thebibliography}
\end{document}